\documentclass{llncs}

\usepackage{amssymb,amsfonts,mathrsfs,epsfig}

\newcommand{\limplies}{\rightarrow}
\newcommand{\liff}{\leftrightarrow}
\newcommand{\fa}[1]{\forall #1 \;} 

\newcommand{\denotes}{\mathord{\downarrow}}
\newcommand{\myrule}{\noindent \rule{\textwidth}{0.3mm}}

\newcommand{\na}[1]{\mathit{#1}}    
\newcommand{\fn}[1]{\mathit{#1}}    

\newcommand{\lip}{\langle}
\newcommand{\rip}{\rangle}

\newcommand{\st}{ \; | \; } 
\newcommand{\ph}{\varphi}

\newcommand{\seq}[1]{\left\langle #1 \right\rangle}
\newcommand{\set}[1]{\left\lbrace #1 \right\rbrace}
\newcommand{\less}{\backslash}

\title{A language for mathematical\\ knowledge management}

\author{Steven Kieffer\inst{1} \and Jeremy Avigad \inst{2} \and 
  Harvey Friedman\inst{3}\thanks{Work by Avigad and Friedman partially
  supported by NSF grant DMS-0700174.}}

\institute{Simon Fraser University \and Carnegie Mellon University \and 
  Ohio State University}

\begin{document}

\maketitle

\begin{abstract}
  We argue that the language of Zermelo Fraenkel set theory with
  definitions and partial functions provides the most promising
  bedrock semantics for communicating and sharing mathematical
  knowledge. We then describe a syntactic sugaring of that language
  that provides a way of writing remarkably readable assertions
  without straying far from the set-theoretic semantics. We illustrate
  with some examples of formalized textbook definitions from
  elementary set theory and point-set topology. We also present
  statistics concerning the complexity of these definitions, under
  various complexity measures.
\end{abstract}

\section{Introduction}

With the growing use of digital means of storing, communicating,
accessing, and manipulating mathematical knowledge, it becomes
important to develop appropriate formal languages for the
representation of such knowledge. But the scope of
``mathematical knowledge'' is broad, and the meaning of the word
``appropriate'' will vary according to the application. At the
extremes, there are competing desiderata:
\begin{itemize}
\item At the \emph{foundational level}, one wants a small and simple
  syntax, and a precise specification of its semantics. In particular,
  one wants a specification as to which inferences are valid.
\item At the \emph{human level}, one wants to have mathematical
  languages that are as easy to read and understand as ordinary
  mathematical texts, yet also admit a precise interpretation to the
  foundational level.
\end{itemize}
For ordinary working mathematicians, the foundational interpretation
is largely irrelevant, but some sort of formal semantics is necessary
if the information encoded in mathematical texts is to be used and
manipulated at the formal level. Of course, one solution is simply to
pair each informal mathematical assertion with a formal translation,
but then there is the problem of obtaining the formal translations and
ensuring that they match the intention of the informal text. As a
result, it is more promising to use semi-structured languages that
integrate features of both the foundational and human levels.  This
results in a smooth spectrum of languages in between the two extremes.
At intermediate ``expert user'' levels, one may want a language whose
structure is close to that of the underlying foundational framework,
yet is as humanly readable as possible.

To complicate matters, there are features of mathematical knowledge
that are not captured at the level of assertions: mathematical
language is used to communicate definitions, theorems, proofs,
algorithms, and problems, among other things. At the level of a
mathematical theory, language is also used to communicate
relationships between these different types of data. The formal
information that is relevant will vary depending on the application
one has in mind, be it database access and search, theorem proving,
formal verification, etc.

Here we will be primarily concerned with mathematical assertions as
they are used to state definitions and theorems.\footnote{In passing,
  we note that computational proof assistants like Mizar
  \cite{rudniki:92}, HOL~\cite{gordon:melham:93},
  Isabelle~\cite{nipkow:et:al:02}, Coq~\cite{bertot:casteran:04} and
  HOL light~\cite{harrison:96} all provide languages that can be used
  to describe mathematical proofs. Of these, the Mizar and
  Isabelle/Isar languages model human proof languages most closely.
  The Isar effort~\cite{wenzel:07} shows that the proof language is
  somewhat orthogonal to the assertion language; that is, Isar can be
  instantiated to various foundational frameworks, subject only to
  minor constraints.} If one is looking for a foundational
framework that is robust enough to subsume those used by most systems
of MKM, it is hard to beat the language of set theory: we know of no
foundational system other than Quine's New Foundations that cannot be
interpreted in the language of set theory in such a way that
inferences are reduced to inferences in Zermelo-Fraenkel set theory
with the axiom of choice ($\na{ZFC}$), or some plausible extension
(say, with large universes of sets). To be clear, we are not denying
the importance of other frameworks for more specific purposes.  For
example, the theory of real closed fields is appropriate to
representing many constraint problems, and constructive frameworks are
better suited to certain forms of algorithmic reasoning. It is also
important to find ways of sharing the additional information that
comes with the use of these more restricted frameworks. We are simply
singling out set theory as a unifying framework for expressing what
assertions in the various local frameworks have in common.

We extend the foundational framework in two ways. First, we allow for
explicit definitions of new predicates and functions on the universe
of sets. And, second, we allow function symbols to denote functions
that are only partially defined, using a logic of partial terms. We
call the resulting formal system $\na{DZFC}$. As we observe in
Section~\ref{dzfc:section}, this system is easily shown to be
conservative over $\na{ZFC}$.  We argue that these extensions are
\emph{not} just a matter of syntactic sugar, but, rather, are
essential to adequate representation of the mathematical data: there
is a difference between assertions using defined terms and their
expanded versions, and, in mathematical terms, $1/0$ really is an
undefined quantity. Thus $\na{DZFC}$ is our proposal for a
foundational language and its semantics.

Our main goal here is to show that the distance between this
foundational level and ordinary mathematical text is not as far as is
commonly supposed, by presenting a syntactically-sugared version of
set theory, $\na{PST}$, that is simultaneously close to both. On the
one hand, we show that our language is easily parsed and translated to
$\na{DZFC}$. On the other hand, by automatically replacing symbolic
expressions with user-provided natural language equivalents, we obtain
output that is humanly readable, and, although not exactly literary,
recognizably faithful to the original mathematical texts. 

We support this last claim with examples from Suppes's \emph{Axiomatic
  set theory} \cite{suppes} and Munkres's \emph{Topology}
\cite{munkres}. In each case, we present our formal input with both
$\na{DZFC}$ and our natural language translations.
Indeed, the appendices to
Kieffer~\cite{kieffer:07} provide a corpus of 341 definitions, taken
from Chapters 2--6 of Suppes and Sections 12--38 of Munkres. Examples
of the natural language translations can be found in Appendix B,
below. These examples show that $\na{PST}$ offers a promising target
semantics for mathematical markup languages, like OMDoc
\cite{kohlhase:06}.

To illustrate the utility of $\na{PST}$, we describe two pieces of
software that take advantage of both the formal structure of the
definitions and their proximity to the informal text.  First, we
describe statistical studies of the complexity of definitions in our
corpus, measured in various ways. Our analysis shows, not
surprisingly, that expanding definitions to the pure language of set
theory yields formulas that are huge. Perhaps more surprisingly,
quantifier complexity of definitions remains remarkably low, even when
they are expanded to $\na{DZFC}$. We also describe software that makes
it possible to explore definitional dependencies, expanding and
compressing nodes via a graphical interface. To be sure, data like
this can be mined from contemporary formal verification
efforts.\footnote{See, for example, the MPTP challenges,
  http://www.cs.miami.edu/$\sim$tptp/MPTPChallenge/} But mathematical
developments are often changed significantly in the process of
formalization; what distinguishes the data presented here is the
extent to which it faithfully represents the informal texts it is
supposed to model.

Our ``user-friendly'' version of set theory is based on
Friedman~\cite{friedman:unp:05}; see also an earlier version in
Friedman~\cite{friedman:flagg:90}. Most of the work described here,
including the implementation of the parser, the entering of the data
from Suppes's and Munkres's books, and associated software,
constitute Kieffer's MS thesis~\cite{kieffer:07}, written under
Avigad's supervision. The thesis and code described here, as well as
additional samples of the natural language translations, can be found
via Avigad's web page.\footnote{Specifically, see
  http://www.andrew.cmu.edu/user/avigad/Papers/mkm/.}

\section{$\na{ZFC}$ with definitions and partial terms}
\label{dzfc:section}

It is widely acknowledged that Zermelo-Fraenkel axiomatic set theory
with the axiom of choice, $\na{ZFC}$, is robust enough to accommodate
ordinary mathematical arguments in a straightforward way. The most
notable exceptions are category-theoretic arguments which rely on the
existence of large universes with suitable closure properties; but
these can be formalized in extensions of $\na{ZFC}$ with suitable
large cardinal axioms, or by restricting the closure properties of the
universes in question. 

In this section, we describe a conservative extension $\na{DZFC}$ of
$\na{ZFC}$. This theory incorporates two features that allow for a
more direct and natural mathematical modeling:
\begin{itemize}
\item it accommodates partially defined functions, and hence
  undefined terms; and
\item it allows the introduction of new function and predicate
  symbols to stand for explicitly defined functions and predicates.
\end{itemize}
We describe each of these extensions, in turn. 

To start with, $\na{DZFC}$ is based on a free logic, with a special
predicate $E(t)$. This is usually written $t \denotes$, and can be
read ``$t$ is defined'' or ``$t$ denotes.'' The axioms governing the
terms are presented as the ``logic of partial terms'' in Beeson
\cite{beeson:85}, $E^+$ logic in Troesltra and Schwichtenberg
\cite{troelstra:schwichtenberg:00}; see also the very helpful
explanation and overview in Feferman~\cite{feferman:95}. The basic idea
is that variables in the language range over objects in the intended
domain (in our case, sets), but, as function symbols may denote
partial functions, some terms fail to denote. So, for example, the
axioms for universal instantiation are given by $\fa x \ph(x) \land t
\denotes \limplies \ph(t)$. The basic relation symbols of
$\na{ZFC}$, which we take to be $\in$ and $=$, are assumed only to
hold between terms that denote; thus we have axioms $s \in t \limplies
s \denotes \land t \denotes$ and $s = t \limplies s \denotes
\land t \denotes$. Partial equality $s \simeq t$ is defined as usual
by the axiom $s \simeq t \liff (s \denotes \lor t \denotes
\limplies s = t)$. 

Next, the syntax of ordinary set theory is extended to include
definition descriptions, \emph{\`a la} Russell. Formally, for each
formula $\ph(x)$, the expression $(\iota x) \ph(x)$ is a term whose
free variables are just those of $\ph$, other than $x$. These terms
are governed by the axioms
\[
y = (\iota x) \ph(x) \liff \fa z(\ph(z) \liff z = y).
\]
Thus in $\na{DZFC}$ one can show that $(\iota x) \ph(x)$ is defined if
and only if there is a unique $y$ satisfying $\ph(y)$, in which case,
$(\iota x) \ph(x)$ is equal to that $y$. 

Finally, one is allowed to introduce new function symbols and relation
symbols to abbreviate formulas and terms. That is, for each formula
$\ph(x,\bar y)$, one can introduce a new function symbol $f(\bar y)$
with the axiom
\[
f(\bar y) \simeq (\iota x) \ph(x,\bar y),
\]
and for every formula $\psi(\bar y)$ one can introduce a new relation
symbol $R(\bar y)$ with the axiom
\[
R(\bar y) \liff \psi(\bar y).
\]
It is not hard to show that adding the usual axioms of set theory to
this framework yields a conservative extension:

\begin{theorem}
$\na{DZFC}$ is a conservative extension of $\na{ZFC}$.
\end{theorem}

The proof amounts to an interpretation of partial functions and
elimination of definitions that is by now standard; details can be
found in \cite{troelstra:schwichtenberg:00,kieffer:07}. Note, however,
that the usual method of eliminating defined function symbols and
relation symbols by replacing them by their definiens can result in an
exponential increase in length.

\section{The language of practical set theory, $\na{PST}$}
\label{PST_section}

We now describe a more flexible language, \emph{Practical set theory},
or $\na{PST}$, designed by Friedman. This language has two key
features:
\begin{itemize}
\item The language incorporates a healthy amount of syntactic sugar,
  making it possible to express ordinary mathematical definitions and
  assertions in a natural way.
\item The language is easily and efficiently translatable to
  $\na{DZFC}$. 
\end{itemize}
In this section we describe some of the features of $\na{PST}$ and
the translation to $\na{DZFC}$. A full and precise specification of
the $\na{PST}$ and its $\na{DZFC}$ semantics can be found in
\cite{kieffer:07,friedman:unp:05}, where it was called the
\emph{Language of Proofless Text}, or $\na{LPT}$.  The claims of
naturality will be supported with examples in the next section and in
Appendix B.

The starting point for $\na{PST}$ is the usual syntax of first-order
logic. We adopt conventions to distinguish between variables, defined
functions, and relations; application of a defined relation $\fn{REL}$
to terms $t_1,\ldots,t_k$ is written with square brackets
$\fn{REL}[t_1,\ldots,t_k]$, while application of a defined function
$\fn{Fun}$ is written with parentheses,
$\fn{Fun}(t_1,\ldots,t_k)$. The usual language of first-order logic is
augmented with a significant amount of ``syntactic sugar,'' to make
the expression of mathematical notions as convenient as
possible. These include the following.

\bigskip

\noindent \emph{Function application for sets.} Any term may be used
as though it were a function, of any arity (including ``infix''). For
example, one may quantify a variable $f$, and then proceed to use it
as though it were a function. In $\na{PST}$, $f(x)$ denotes the unique
$u$ such that the ordered pair $\lip x, u \rip$ is in $f$,
assuming there is such $u$. The following definition of the unary
predicate {\tt FCN} therefore asserts that $f$ is a function if it is
a set of ordered pairs $\lip x, u \rip$ in which no $x$ occurs
more than once as the first component of a pair.

\medskip

\myrule

\noindent DEFINITION FS.2.58: 1-ary relation $\mathop{\mathtt{FCN}}$.
$\mathop{\mathtt{FCN}}[f] \leftrightarrow f = \lbrace \seq{x,y} : f(x) = y
\rbrace $.

\myrule

\bigskip

\noindent \emph{Finite sets and tuples.} In the previous example, we saw a finite tuple; namely, the ordered pair $\seq{x,y}$. Tuples of any finite length are terms in $\na{PST}$.

A finite set can be denoted by simply listing all of its elements. For example, in defining the Wiener-Kuratowski ordered pair, we may use the term $\{\{a\},\{a,b\}\}$.

\bigskip

\noindent \emph{Set-builder notation.} The example above illustrates
the use of set-builder notation. In $\na{PST}$, the term $\{t:\ph\}$
denotes the set of all values of $t(x_1,\ldots,x_n)$, where the
variables $x_1, \ldots, x_n$ occurring in $t$ range over tuples
satisfying $\ph(x_1,\ldots,x_n)$. Note that this involves an essential
use of partiality; for example, in the intended semantics, the term
$\{ x : x = x \}$ is undefined.

Suppose we wish to define $\mathtt{Image}(f)$ to be the
set of all $f(x)$ such that $x \in \mathtt{Dom}(f)$. The expression
\[ \mathtt{Image}(f) \simeq \set{f(x) : x \in \mathtt{Dom}(f)} \] is
not what we want, because $f$ on the right-hand side is taken to be a
bound variable ranging over the universe of sets. Instead, $\na{PST}$
has us write
\[ \mathtt{Image}(f) \simeq \set{f(x) : x \in \mathtt{Dom}(f),
   \mbox{ $f$ fixed} } \]
to indicate that the expression depends on a fixed value of $f$. 

\bigskip

\noindent \emph{Defined function symbols.} We use an exclamation mark
in place of Russell's $\iota$ as a definite description operator.
It is used in the next example, where we define an infix function,
$+_{\mathbb{Q}}$, for addition on the rational numbers. Every infix function
is given a \emph{precedence} number, for use in determining order of
operations.

\medskip

\myrule

\noindent DEFINITION FS.5.25: Infix function $\mathop{\mathtt{+_{\mathbb{Q}}}}$. $x
\mathop{\mathtt{+_{\mathbb{Q}}}} y \simeq (! z)(x,y,z \mathop{\mathtt{\in}}
\mathop{\mathtt{\mathbb{Q}}} \wedge (\exists a,b,c)$

\noindent $(a \mathop{\mathtt{\in}} x \wedge b
\mathop{\mathtt{\in}} y \wedge c \mathop{\mathtt{\in}} z \wedge a +_{SUB} b =
c))$. Precedence 40.

\myrule

\medskip
A definition may be composed of any number of ``If ... then ...'' clauses, and may end with one ``Otherwise ...'' clause, which allows definition by cases, as in the example below. In this example the `Otherwise' clause introduces a condition under which the function is undefined. For this we use the predicate $\uparrow$, and this allows for the definition of partial functions.

\medskip

\myrule

\noindent DEFINITION FS.2.3: 1-ary function $\mathop{\mathtt{Dom}}$. If
$\mathop{\mathtt{BR}}[R]$ then $\mathop{\mathtt{Dom}}(R) \simeq \lbrace x :
(\exists y)(x \mbox{ $R$ } y) \rbrace $. Otherwise
$\mathop{\mathtt{Dom}}(R)\mathord{\uparrow}$.

\myrule

\bigskip

\noindent \emph{Defined relation symbols.} As with functions, we may define infix relations, as in the definition of $<$ on the rational numbers, below.
\medskip

\myrule

\noindent DEFINITION FS.5.24: Infix relation
$\mathop{\mathtt{<_{\mathbb{Q}}}}$. $x
\mathop{\mathtt{<_{\mathbb{Q}}}} y \leftrightarrow (\exists z,w)(x,y
\mathop{\mathtt{\in}} \mathop{\mathtt{\mathbb{Q}}} \wedge z
\mathop{\mathtt{\in}} x \wedge w \mathop{\mathtt{\in}} y \wedge z
\mathop{\mathtt{<_{SUB}}} w)$.

\myrule

\bigskip

\noindent \emph{Lambda notation.} $\na{PST}$ includes a lambda operator
which can be used to bind variables and thereby denote functions. In
the example below, we define a binary function called {\tt Cartespow}
(for ``Cartesian power''). This function maps a pair of sets $A$, $B$
to the set $A^B$; i.e., a product of $B$-many copies of $A$. The
definition relies on a previously defined function, {\tt Cartesprod}
(for ``Cartesian product''), a binary function taking a map $f$ and a
set $C$ to the product over $c \in C$ of the sets $f(c)$. The
definition of {\tt Cartespow} uses lambda abstraction to define the
constant function $b \mapsto A$ on the fly, to serve as the first
argument to {\tt Cartesprod}.

\medskip

\myrule

\noindent DEFINITION MunkTop.19.2.5: 2-ary function
$\mathop{\mathtt{Cartespow}}$. $\mathop{\mathtt{Cartespow}}(A,B) \simeq
\mathop{\mathtt{Cartesprod}}((\lambda b \mathop{\mathtt{\in}} B)(A),B)$.

\myrule

\bigskip

\noindent \emph{Infix relation chains.} Infix relations may be chained
together in the usual way, as with the $<_{\mathtt{\mathbb{R}}}$
relation in the example below.

\medskip

\myrule

\noindent DEFINITION MunkTop.13.3.a.basis: 0-ary function
$\mathop{\mathtt{Stdrealtopbasis}}$.

\noindent $\mathop{\mathtt{Stdrealtopbasis}} \simeq
\lbrace U \mathop{\mathtt{\subseteq}} \mathop{\mathtt{\mathbb{R}}} : (\exists
a,b \mathop{\mathtt{\in}} \mathop{\mathtt{\mathbb{R}}})(U = \lbrace x
\mathop{\mathtt{\in}} \mathop{\mathtt{\mathbb{R}}} : a \mathop{\mathtt{<_{\mathbb{R}}}} x
\mathop{\mathtt{<_{\mathbb{R}}}} b \rbrace ) \rbrace $.

\myrule

\bigskip

\noindent \emph{Bounded quantifiers.} Quantified variables and
variables used in set-builder notation may be bounded by any infix
relation, as in the example above.

\bigskip

The translation from $\na{PST}$ to $\na{DZFC}$ is not difficult. Since
our grammar for $\na{PST}$ is not LL, we used the ACCENT compiler-compiler
\footnote{http://accent.compilertools.net/}, which implements
Earley's algorithm. The latter can parse any context-free grammar in
cubic time, and runs in quadratic time when the grammar is unambiguous
\cite{aho:ullman}.

Appendix A contains a number of examples of $\na{PST}$ definitions,
together with their translations to $\na{DZFC}$. In each case, we
present the $\na{DZFC}$ input, a \LaTeX{} representation of that input
generated by the parser, and the translation to $\na{DZFC}$. A much
larger corpus of examples --- 183 definitions from Suppes's
\emph{Axiomatic Set Theory} \cite{suppes} and 148 definitions from
Munkres's \emph{Topology} \cite{munkres} --- can be found in
\cite{kieffer:07}. In practice, the translation took at most a few
seconds to process a file containing a dozen large
definitions. Comparing the (\LaTeX{}) $\na{DZFC}$ output with the (\LaTeX{}
version of the) $\na{PST}$ input yields a factor of about $0.91$,
which is to say, the $\na{DZFC}$ translations are actually slightly
shorter.

\section{Natural language output}
\label{NLoutputSection}

The examples of $\na{PST}$ input in the last section are readable, but
not attractive. It is hard to remember meaning of symbols ``BR'' or
``TOPSP''; it would help to have phrases like ``is a binary relation''
or ``is a topological space.'' In fact, even for logical connectives
like $\land$, natural language equivalents like ``and'' are generally
easier to read. In an ordinary mathematical language text, however,
words are not always favored over symbols. For example, defined
functions are usually given symbols: $\fn{gcd}(x,y)$ instead of ``the
greatest common divisor of $x$ and $y$.'' Binary relations like $=$
and $<$ are usually preferred to ``equal to'' and ``less than.'' On
the other hand, unary relations often represent concepts that are
expanded to words, as shown by the examples above.

In light of these observations, we chose to output natural language
equivalents for the connectives, and allow the user to input natural
language equivalents for defined symbols. For example, with the entry
\begin{center}
\begin{verbatim}
TOPSP:2@
  reln:$(#^0,#^1)$ is a %e?topological space%ee?@
  negn:$(#^0,#^1)$ is not a topological space@
  plur:%$(#^0,#^1)$% are topological spaces@
  nplu:%$(#^0,#^1)$% are not topological spaces@@
\end{verbatim}
\end{center}
the user can specify the natural language that should be used in place of the {\tt TOPSP} relation.

In some cases, either symbols or a natural language equivalent can be
used, as in $\{ x \in \Bbb N \st \ldots \}$ or ``the set of $x \in
\Bbb N$ such that \ldots.'' It is usually awkward to have natural
language occur as a subterm of a symbolic expression; for example,
consider ``$1 + \mbox{the greatest common divisor of $x$ and $y$}$.''
Thus we incorporate a monotonicity rule: once a subterm of a term has
been expanded to natural language, natural language versions are
favored from then on. This choice yields, for example, $\{ x \in
\Bbb N \st a < x < b \}$, but also ``the set of $x$ in $\Bbb N$
such that $a < x < b$ and $x$ is even.''

Accordingly, the user supplies two clauses for a defined function or relation for which symbols are preferred over words:
\begin{verbatim}
\wp:1@
  symb:$\wp(#^0)$@
  word:the power set of #0@@
\end{verbatim}
whereas if words are the desired default then just one clause is needed:
\begin{verbatim}
Stdrealtop:0@
  word:the standard topology on $\mathbb{R}$@@
\end{verbatim}

Appendix B provides examples of natural language output. We emphasize
that these were generated directly from the $\na{PST}$ input, using
the additional natural language data, supplied by the user, described
above. Although the definitions are not exactly literary, they are
surprisingly readable, and close to ordinary mathematical text. It is
certainly the case that additional heuristics could be used to render
the output more attractive, and additional markup from the user would
result in improvements. In other words, there is a lot more that can
be done along these lines; our claim here is only that $\na{PST}$
offers an auspicious start.

\section{Exploring definitions}

Among the benefits of having a database of definitions is the ability to explore those definitions interactively. We designed two simple programs with which to demonstrate some of the possibilities.

Our first program allows the interactive display and manipulation of
directed acyclic graphs (dags) of conceptual dependencies, as depicted
in Figure 1.

\begin{figure}
\begin{center}
\label{kmap_screenshot}
\epsfig{figure=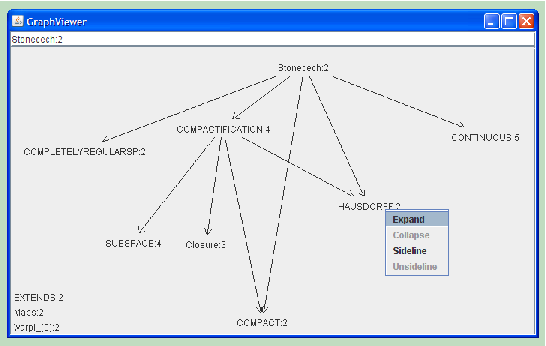, width=\textwidth}
\caption[]{Exploring the definition dag for the Stone-\v Cech compactification.}
\end{center}
\end{figure}

With a second program we gathered statistics on these graphs.
Associated to each definition is the dag of all definitions on which
it depends; by the \emph{size} of this dag we mean the number of
vertices, and by the \emph{depth} of this dag we mean the length of
its longest directed path. Table 1 shows the maximum and mean values
for all definitions in our database.

\begin{table}
\label{dag_data}
\begin{center}
\caption{Max and mean dag sizes and depths}
\begin{tabular}{|r|r|c|c|}
\hline
& & Max & Mean \\\hline
All & Depth & 32 & 10.77 \\
  & Size & 110 & 29.56 \\\hline
Suppes & Depth & 26 & 10.09 \\
  & Size & 77 & 25.91 \\\hline
Munkres & Depth & 32 & 12.25 \\
  & Size & 110 & 36.01 \\
\hline
\end{tabular}
\end{center}
\end{table}
Additional statistics, including data on the quantifier complexity of
definitions in our corpus, can be found in Appendix C. 

\section{Conclusions}

We have argued that one should adopt a language close to definitional
set theory as a uniform language to support communication and exchange
of mathematical results. The particular language we describe here,
\emph{Practical set theory}, fares well in that regard: it
is easy and natural to work with, providing a high-degree of
readability while remaining close to a clear foundational semantics.

\section*{Appendix A: Examples of $\na{PST}$ input and $\na{DZFC}$
  translations} 
\label{PST_DZFC_appendix}

We consider a few examples of formal definitions, highlighting the
naturality of $\na{PST}$ over $\na{DZFC}$. (The $\varpi_0$ function
appearing in the $\na{DZFC}$ translations is a function defined to
take $(a,b)$ to the Wiener-Kuratowski ordered pair $\{ \{a\}, \{ a,b
\} \}$.)

\smallskip
\myrule

\medskip

\noindent {\bf Example 1.} Here the description operator is used in
$\na{PST}$ to bind an ordered pair, so that we are able to refer to
``the unique ordered pair $\left<Y,T'\right>$ such that....'' This
translates to a much clumsier expression in $\na{DZFC}$, requiring two
additional bound variables.

\medskip
\noindent \emph{$\na{PST}$ input:}

\vspace{-2mm}

\begin{verbatim}
DEFINITION MunkTop.29.4: 2-ary function Oneptcompactification.
If TOPSP[X,T] then Oneptcompactification(X,T) \simeq
(!<Y,T'>)(
  COMPACTIFICATION[Y,T',X,T] \wedge Y \less X \approx_{C} 1_{N}
).
\end{verbatim}

\noindent \emph{$\na{PST}$ rendered in \LaTeX{}:}

\medskip
\noindent DEFINITION MunkTop.29.4: 2-ary function
$\mathop{\mathtt{Oneptcompactification}}$. If \hfil\break $\mathop{\mathtt{TOPSP}}[X,T]$
then $\mathop{\mathtt{Oneptcompactification}}(X,T) \simeq (!
\seq{Y,T'})$ \hfil\break $(\mathop{\mathtt{COMPACTIFICATION}}[Y,T',X,T] \wedge Y \less X
\mathop{\mathtt{\approx_{C}}} \mathop{\mathtt{1_{N}}})$.

\bigskip
\noindent \emph{$\na{DZFC}$ translation:}

\medskip
\noindent $\mathop{\mathtt{Oneptcompactification}}(X,T) \simeq (\iota
y_{0})(\mathop{\mathtt{TOPSP}}[X,T] \wedge y_{0} \simeq (\iota x_{0})(\exists
Y,T')(x_{0} = \mathop{\mathtt{\varpi_{0}}}(Y,T') \wedge
(\mathop{\mathtt{COMPACTIFICATION}}[Y,T',X,T] \wedge
\mathop{\mathtt{\approx_{C}}}[\mathop{\mathtt{\less}}(Y,X),\mathop{\mathtt{1_{N}}}])))$

\myrule

\medskip
\noindent {\bf Example 2.} Next observe what happens in $\na{DZFC}$,
where we cannot match the brevity of expression used in our definition
of the {\tt FCN[f]} predicate in $\na{PST}$ (which says that {\tt f}
is a function).

\bigskip
\noindent \emph{$\na{PST}$ input:}

\vspace{-2mm}

\begin{verbatim}
DEFINITION FS.2.58: 1-ary relation FCN. FCN[f] \iff
f = {<x,y> : f(x) = y}.
\end{verbatim}

\noindent  \emph{$\na{PST}$ rendered in \LaTeX{}:}

\medskip
\noindent DEFINITION FS.2.58: 1-ary relation $\mathop{\mathtt{FCN}}$.
$\mathop{\mathtt{FCN}}[f] \leftrightarrow f = \lbrace \seq{x,y} : f(x) = y
\rbrace $.

\bigskip

\noindent \emph{$\na{DZFC}$ translation:}

\medskip
\noindent $\mathop{\mathtt{FCN}}[f] \leftrightarrow f = (\iota z_{0})(\forall y_{0})(y_{0}
\in z_{0} \leftrightarrow (\exists x,y)(y_{0} =
\mathop{\mathtt{\varpi_{0}}}(x,y) \wedge ((\iota
x_{0})(\mathop{\mathtt{\varpi_{0}}}(x,x_{0}) \in f) = y)))$

\myrule

\medskip

\noindent {\bf Example 3.} Here we see how important the lambda operator is:

\medskip
\noindent \emph{$\na{PST}$ input:}

\vspace{-2mm}

\begin{verbatim}
DEFINITION MunkTop.19.2.5: 2-ary function Cartespow. Cartespow(A,B)
\simeq Cartesprod((\lambda b \in B)(A),B).
\end{verbatim}

\noindent \emph{$\na{PST}$ rendered in \LaTeX{}:}

\medskip
\noindent DEFINITION MunkTop.19.2.5: 2-ary function
$\mathop{\mathtt{Cartespow}}$. $\mathop{\mathtt{Cartespow}}(A,B) \simeq
\mathop{\mathtt{Cartesprod}}((\lambda b \mathop{\mathtt{\in}} B)(A),B)$.

\bigskip
\noindent \emph{$\na{DZFC}$ translation:}

\medskip
\noindent $\mathop{\mathtt{Cartespow}}(A,B) \simeq \mathop{\mathtt{Cartesprod}}((\iota
z_{0})(\forall y_{0})(y_{0} \in z_{0} \leftrightarrow (\exists b, x_{0})(y_{0} =
\mathop{\mathtt{\varpi_{0}}}(b,x_{0}) \wedge x_{0} = (A) \wedge b \in B)),B)$

\myrule

\section*{Appendix B: Examples of the natural language translations}

In some cases our natural language generating program {\tt pst2nl}
produces output that is quite close to what a human being might
write. For example, from the following $\na{PST}$ input,

\medskip

DEFINITION MunkTop.13.2: 2-ary function
$\mathop{\mathtt{Basisgentop}}$. If
$\mathop{\mathtt{TOPBASIS}}[\mathscr{B},X]$ then
$\mathop{\mathtt{Basisgentop}}(\mathscr{B},X) \simeq (! \mathscr{T}
\mathop{\mathtt{\subseteq}} \mathop{\mathtt{\wp}}(X))((\forall U
\mathop{\mathtt{\subseteq}} X)(U \mathop{\mathtt{\in}} \mathscr{T}
\leftrightarrow (\forall x \mathop{\mathtt{\in}} U)(\exists B
\mathop{\mathtt{\in}} \mathscr{B})$

\noindent $(x \mathop{\mathtt{\in}} B \wedge B
\mathop{\mathtt{\subseteq}} U)))$.

\medskip
\noindent we get the following NL (natural language) output:

\medskip

\vbox{
\myrule

{\bf Definition:} If $\mathscr{B}$ is a basis for a topology on $X$
then \emph{the topology on $X$ generated by $\mathscr{B}$} is the
unique $\mathscr{T}$ $\subseteq$ $\wp(X)$ such that for every $U$
$\subseteq$ $X$, $U$ $\in$ $\mathscr{T}$ if and only if for every $x$
$\in$ $U$, there exists $B$ $\in$ $\mathscr{B}$ such that $x$ $\in$
$B$ and $B$ $\subseteq$ $U$.

\myrule
}

Indeed, this is not substantially different from the original text in
Munkres \cite{munkres}, page 78. After defining what it means for
$\mathscr{B}$ to be a basis, Munkres says,

\begin{quote}
  If $\mathscr{B}$ satisfies these two conditions, then we define the
  \emph{topology $\mathscr{T}$ generated by} $\mathscr{B}$ as follows:
  A subset $U$ of $X$ is said to be open in $X$ (that is, to be an
  element of $\mathscr{T}$) if for each $x \in U$, there is a basis
  element $B \in \mathscr{B}$ such that $x \in B$ and $B \subset
  U$. Note that each basis element is itself an element of
  $\mathscr{T}$.
\end{quote}

\medskip What is more common is that the output of {\tt pst2nl} reads
nicely except for a ``run-on'' sound, resulting from insufficient
punctuation. For example:

\medskip

\vbox{
\myrule

{\bf Definition:} If $R$ is a strong simple order on $X$ then
\emph{the basis for the order topology on $(X,R)$} is the set of $U$
such that there exist $a$, $b$ $\in$ $X$ such that $U$ $=$ $(a,b)$ or
$a$ is a first element in $X$ and $U$ $=$ $[a,b)$ or $b$ is a last
element in $X$ and $U$ $=$ $(a,b]$.

\myrule
}

In Munkres, page 84, all of this information is spread out over a
numbered list:
\begin{quote}
  \noindent {\bf Definition.} Let $X$ be a set with a simple order
  relation; assume $X$ has more than one element. Let $\mathscr{B}$ be
  the collection of all sets of the following types:
\begin{enumerate}
\item All open intervals $(a,b)$ in $X$.

\item All intervals of the form $[a_0,b)$, where $a_0$ is the smallest
  element (if any) of $X$.

\item All intervals of the form $(a,b_0]$, where $b_0$ is the largest
  element (if any) of $X$.
\end{enumerate}
The collection $\mathscr{B}$ is a basis for a topology on $X$, which
is called the \emph{order topology}.
\end{quote}

\medskip
Heuristics, combined with additional user markup, could eventually be
incorporated to help improve the flow and punctuation of the
translations. We have implemented one easy improvement already,
whereby adjacent assertions of a common predicate are combined into a
single assertion using plural form. Thus, from the $\na{PST}$ input,

\medskip

DEFINITION MunkTop.12.4.a: 3-ary relation
$\mathop{\mathtt{FINERTOP}}$. If
$\mathop{\mathtt{TOPSP}}[X,\mathscr{T}] \wedge
\mathop{\mathtt{TOPSP}}[X,\mathscr{T}']$ then
$\mathop{\mathtt{FINERTOP}}[\mathscr{T}',\mathscr{T},X]
\leftrightarrow \mathscr{T}' \mathop{\mathtt{\supseteq}} \mathscr{T}$.

\medskip
\noindent we obtain:

\medskip

\vbox{
\myrule

{\bf Definition:} If $(X,\mathscr{T})$ and $(X,\mathscr{T}')$ are
topological spaces then $\mathscr{T}'$ is \emph{finer} than
$\mathscr{T}$ on $X$ if and only if $\mathscr{T}'$ $\supseteq$
$\mathscr{T}$.

\myrule
}

This time Munkres is able to make several definitions in a single
paragraph, and can abbreviate a more complex logical locution with
the phrase ``respective situations.''

\begin{quote}
  \noindent {\bf Definition.} Suppose that $\mathscr{T}$ and
  $\mathscr{T}'$ are two topologies on a given set $X$. If
  $\mathscr{T}' \supset \mathscr{T}$, we say that $\mathscr{T}'$ is
  \emph{finer} than $\mathscr{T}$; if $\mathscr{T}'$ \emph{properly}
  contains $\mathscr{T}$, we say that $\mathscr{T}'$ is \emph{strictly
    finer} than $\mathscr{T}$. We also say that $\mathscr{T}$ is
  \emph{coarser} than $\mathscr{T}'$, or \emph{strictly coarser}, in
  these two respective situations. We say $\mathscr{T}$ is
  \emph{comparable} with $\mathscr{T}'$ if either $\mathscr{T}'
  \supset \mathscr{T}$ or $\mathscr{T} \supset
  \mathscr{T}'$.
\end{quote}

\medskip
We consider a final example,

\medskip

\noindent DEFINITION MunkTop.13.3.c: 0-ary function
$\mathop{\mathtt{Krealtop}}$. $\mathop{\mathtt{Krealtop}} \simeq$
\hfil\break
$\mathop{\mathtt{Basisgentop}}(\mathop{\mathtt{Stdrealtopbasis}} \cup
\lbrace V \mathop{\mathtt{\subseteq}} \mathop{\mathtt{\mathbb{R}}} :
(\exists W \mathop{\mathtt{\in}} \mathop{\mathtt{Stdrealtopbasis}})$
\hfil\break $(V = W \less \lbrace
\mathop{\mathtt{Incl_{FrR}}}(\mathop{\mathtt{1_{N}}} / n) : n
\mathop{\mathtt{\in}} \mathop{\mathtt{\mathbb{N}}} \rbrace ) \rbrace
,\mathop{\mathtt{\mathbb{R}}})$.

\medskip
\noindent for which the NL output is as follows:

\medskip

\vbox{
\myrule

{\bf Definition:} \emph{The K-topology on $\mathbb{R}$} is the
topology on $\mathbb{R}$ generated by the standard basis for a
topology on $\mathbb{R}$ union the set of $V$ $\subseteq$ $\mathbb{R}$
such that there exists $W$ in the standard basis for a topology on
$\mathbb{R}$ such that $V$ $=$ $W$ $\backslash$ $\set{1 / n : n \in
  \mathbb{N}}$.  \myrule }

\medskip There are two sets mentioned in this definition: the set of
$V \subseteq \mathbb{R}$ such that ..., and the set of $1/n$ such that
.... According to the ``monotonicity rule'' described in Section
\ref{NLoutputSection}, the latter is rendered in symbols since it has
no subterm in words; the former is rendered in words since its
subterm, ``the standard basis for a topology on $\mathbb{R}$'' has no
symbolic form, and is displayed in words by default.

Another feature of {\tt pst2nl} is apparent in this last example,
where the word ``in'' appears before ``the standard basis....'' We get
this preposition rather than the incorrect phrase ``is in,'' thanks to
the final clause in the user-supplied natural language equivalents for
the $\in$ relation:

\vbox{
\begin{verbatim}
\in:infix@
  symb:#0 $\in$ #1@
  nsym:#0 $\not\in$ #1@
  reln:#0 is %e?in%ee? #1@
  negn:#0 is not in #1@
  plur:%#0% are in #1@
  nplu:%#0% are not in #1@
  prep:#0 in #1@@
\end{verbatim}
}

Finally we note that the user is free to suppress artifacts of
formalization, in the NL output. In the $\na{PST}$ above there is an
inclusion function $\mathtt{Incl_{FrR}}$, and the number $1$ is
subscripted as $1_{\mathtt{N}}$. None of this shows up in the NL
output.

\medskip Comparison with Munkres, page 82, reveals that he is free to
write in a less regimented form than that of our definitions:
\begin{quote}
  Finally, let $K$ denote the set of all numbers of the form $1/n$,
  for $n \in \mathbb{Z}_+$, and let $\mathscr{B}''$ be the collection
  of all open intervals $(a,b)$, along with all sets of the form
  $(a,b) - K$. The topology generated by $\mathscr{B}''$ will be
  called the \emph{$K$-topology} on $\mathbb{R}$.
\end{quote}

\section*{Appendix C: Data on quantifier complexity and length}

Our database of definitions entered in $\na{PST}$ consists of 183
definitions from Suppes's \emph{Axiomatic Set Theory} \cite{suppes}
and 148 definitions from Munkres's \emph{Topology} \cite{munkres}.

\medskip

\noindent \emph{Quantifier complexity data.} For each definition in
our database, we measured quantifier complexity in eight different
ways. In the first place, we considered both alternating quantifier
depth, and non-alternating. Secondly, we considered each definition in
four different states: (1) as given in $\na{PST}$; (2) as translated
into $\na{DZFC}$; (3) the \emph{expanded} version of the $\na{DZFC}$,
that is, with all definienda replaced by their definiens, recursively,
until the process halts; and (4) a \emph{partially expanded} version
of the $\na{DZFC}$ in which certain low-level, foundational definienda
were left unexpanded, namely: the union, intersection, and set
difference operations, the ordered pair, and powerset functions, the
empty set, and the subset and superset relations. The maximum and mean
depths are presented in Table 2.

\begin{table}
\label{qDepths}
\begin{center}
\caption{Max and mean quantifier depths}
\begin{tabular}{|r|r|r|}
\hline
 & Max & Mean \\\hline
$\na{PST}$ & 4 & 0.66 \\
unexpanded $\na{DZFC}$ & 5 & 1.31 \\
fully expanded $\na{DZFC}$ & 1235 & 78.68 \\
partially expanded $\na{DZFC}$ & 552 & 38.54 \\\hline
$\na{PST}$ alternating & 3 & 0.63 \\
unexpanded $\na{DZFC}$ alternating & 5 & 1.18 \\
fully expanded $\na{DZFC}$ alternating & 422 & 36.19 \\
partially expanded $\na{DZFC}$ alternating & 239 & 22.16 \\
\hline
\end{tabular}
\end{center}
\end{table}
It has been said that among actually occurring definitions in
mathematics texts, the maximum alternating quantifier depth is three.
Insofar as $\na{PST}$ comes close to what actually occurs in
textbooks, the maximum alternating depth of 3 tends to
confirm this conjecture.

Note that the maximum depth after translating into $\na{DZFC}$ goes up
to 5. This reflects what we saw in Appendix A, where a definition that
used no quantifiers in $\na{PST}$ turned out to require them after
translation into $\na{DZFC}$.

The maximum depth of 1235 for a fully expanded definition confirms the
necessity of using definitions to package information into manageable
chunks. Meanwhile, the contrast between the total expansion maximum,
and the partial expansion maximum of 552, demonstrates that the
lowest, most foundational definitions, lend quite a bit of this
complexity.

The ratio $78.68/36.19 \approx 2.17$ of the mean fully expanded depth
to the mean fully expanded alternating depth suggests that quantifiers
often occur in runs of two, before alternating, when definitions are
written in pure set theory. The somewhat lower ratio of $38.54/22.16
\approx 1.74$ for the partially expanded cases indicates the extent to
which the lowest-level concepts contribute to this doubling of
consecutive quantifiers.

The mean depth for $\na{PST}$ alternating (again, what comes closest
to what we ordinarily think of as quantifier depth in textbooks) shows
that, while the maximum is three, the most common depths are 0 and
1. The exact number of occurrences are presented in Table 3.

\begin{table}
\label{PSTdepths}
\begin{center}
\caption{Quantifier depth frequencies in $\na{PST}$}
\begin{tabular}{|c|c|c|}
\hline
 & \multicolumn{2}{|c|}{Occurrences} \\\hline
Depth & $\na{PST}$ & $\na{PST}$ alternating \\\hline
0 & 178 & 178 \\
1 & 118 & 120 \\
2 & 30  & 35  \\
3 & 14  & 8   \\
4 & 1   & 0   \\
\hline
\end{tabular}
\end{center}
\end{table}

\noindent \emph{Length data.} As was expected, there is rapid blowup
in the size of definitions when they are expanded. In collecting our
data we set a maximum of $2^{31}-1$ before we stopped counting, and
this maximum was often reached.

In particular, since the development of the real numbers taken from
Suppes \cite{suppes} involves such deep definition trees, any
definition mentioning the real numbers will have enormous expanded
length. For example, the definition of the basis for the standard
topology on the reals (see Section \ref{PST_section}) is just 303
symbols long after initial translation into $\na{DZFC}$, but blows up
to over $2^{31}-1$ symbols after expansion.

The longest definition we formalized from Suppes \cite{suppes} was 526
symbols, and the longest from Munkres \cite{munkres} was 714 symbols.


\end{document}